# Two-Dimensional Dirac Fermions in Thin Films of $Cd_3As_2$

Luca Galletti[1,†,*], Timo Schumann[1,†], Omor F. Shoron[1], Manik Goyal[1], David A. Kealhofer[2], Honggyu Kim[1], and Susanne Stemmer[1,*]

[1]Materials Department, University of California, Santa Barbara, CA 93106-5050, USA.

[2]Department of Physics, University of California, Santa Barbara, CA 93106-9530, USA.

[†] These authors equally contributed to the work

[*] Email: galletti@mrl.ucsb.edu, stemmer@mrl.ucsb.edu

**Abstract**

Two-dimensional states in confined thin films of the three-dimensional Dirac semimetal $Cd_3As_2$ are probed by transport and capacitance measurements under applied magnetic and electric fields. The results establish the two-dimensional Dirac electronic spectrum of these states. We observe signatures of *p*-type conduction in the two-dimensional states as the Fermi level is tuned across their charge neutrality point and the presence of a zero energy Landau level, all of which indicate topologically non-trivial states. The resistance at the charge neutrality point is approximately $h/e^2$ and increases rapidly under the application of a magnetic field. The results open many possibilities for gate-tunable topological devices and for the exploration of novel physics in the zero energy Landau level.

# I. Introduction

Recent observations of the quantum Hall effect [1-3] from confined states in thin films of $Cd_3As_2$, a prototype three-dimensional Dirac semimetal [4-7], open up new opportunities for the experimental study of Dirac fermion physics in a novel class of topological materials. For example, theoretical predictions indicate that three-dimensional Dirac semimetals can be tuned into novel topological states if they are sufficiently thin to open up a band gap in their bulk electronic states. Depending on the boundary conditions, gapless, topologically protected Fermi arc states may exist on the surfaces even when the bulk is gapped [8-13] or the gapped material may transform into a quantum spin Hall (QSH) insulator with one-dimensional edge states [4,12,14-16].

Topological state tuning is challenging in practice. Unintentional nonstoichiometries cause *n*-type doping in $Cd_3As_2$, raising the Fermi level above the Dirac node (charge neutrality point or CNP), which may prevent observation of these states. For example, the Fermi level in some $Cd_3As_2$ single crystals is believed to be located above the Lifshitz transition, causing the transport physics to be dominated by "Kane" electrons, rather than Dirac fermions [17]. Furthermore, boundary conditions, such as an inherent asymmetry between bottom and top surfaces, affect the nature of the electronic states in topological materials in ways that are difficult to predict [12,18]. Modulation of the carrier densities by an electric field would provide essential insights into the physics of Dirac fermions in quantum confined states, allow for tuning the Fermi level to the Dirac node to access different topological (or trivial) states, and is important for potential topological device applications [19,20].

Epitaxial, (112)-oriented thin films of $Cd_3As_2$, which are grown by molecular beam epitaxy (MBE), possess a bulk band gap and robust, gapless, two-dimensional states [1,21].

Studies of quantum oscillations as a function of $Cd_3As_2$ film thickness revealed two-dimensional states even in thicker films, which still showed appreciable parallel conduction through the bulk [21]. These two-dimensional states give rise to an integer quantum Hall effect in the thinnest films [1,21], which possess a sufficiently large band gap in the bulk states to (nearly) eliminate parasitic transport through the bulk at sufficiently low temperatures. While the results are in keeping with theoretical predictions of robust Fermi arc surface states, the observation of the quantum Hall effect is by itself insufficient to establish the topological nature of these two-dimensional states.

Here, we show electrostatic tuning experiments that establish the Dirac physics of these two-dimensional states. At high magnetic fields, quantized Landau levels are tuned by electric fields applied to top gates. Electrostatic tuning through the CNP of the two-dimensional states and the presence of a zero energy Landau level in the quantum capacitance prove that the surface states are topologically non-trivial. In the absence of an applied magnetic field, the resistance at the CNP is approximately $h/e^2$, which raises very interesting questions as to the nature of this state.

## II. Experiment

$Cd_3As_2$ thin films were grown by MBE on (111)B-oriented GaAs substrates, using 130 nm thick (111)GaSb buffer layers, as described elsewhere [22]. $Cd_3As_2$ films grow with the (112) planes of their tetragonal unit cell [23] parallel to the substrate surface. We note that the two bulk Dirac nodes lie along the fourfold symmetry axis, [001], and one expects double Fermi arcs on the (112) surface that connect their projections [8-13]. We investigated $Cd_3As_2$ films of two different thicknesses, 38 nm (sample A) and 43 nm (sample B), slightly thicker than in our previous studies of the quantum Hall effect [1,21]. The samples were diced to fabricate Hall bars with dimensions of 50×50 μm and 100×100 μm both with and without a 30-nm-thick $Al_2O_3$ gate oxides, respectively. The $Al_2O_3$ was deposited by atomic layer deposition (ALD) and gated Hall bar devices were patterned using photolithography techniques and $Ar^+$ ion milling. ALD was carried out at an unusually low temperature (120 °C) to minimize the deterioration of the $Cd_3As_2$. After ALD, an Au/Pt/Ti gate top contact was deposited and patterned using a lift off technique. We observe a systematic increase in the carrier density after gate oxide deposition relative to the same sample without the oxide (see Supplementary Fig. S1 [24]), which we attribute to differences in the surface Fermi level pinning [25]. Furthermore, low temperature ALD dielectrics contain many trap states, whose (dis)charging causes a voltage shift between forward and reverse sweeps of the gate voltage (see Supplementary Fig. S2 [24]). Magnetoresistance measurements were performed in an Oxford Triton dilution refrigerator at a temperature of 40 mK (unless specified otherwise) and magnetic fields ($B$) up to 14 T. Capacitance-voltage (*C-V*) characteristics were measured at 40 mK using a Keithley 4200A parameter analyzer at a measurement frequency of 60 kHz. The quality factor remained high (> 20) over the entire range of applied gate biases.

## III. Results and Discussion

### A. Landau levels and electric field effect

Figure 1(a) shows the magnetoresistances for Hall bar devices on samples A and B without applied gate bias. An incipient quantum Hall effect in the Hall resistance ($R_{xy}$) and Shubnikov – de Haas (SdH) oscillations in the longitudinal resistance ($R_{xx}$) can be detected. The two-dimensional carrier densities ($n_{SdH}$) of the carriers in the surface states, which give rise to the quantum oscillations, are extracted from the fan diagram [Fig. 1(b)] and correspond to $n_{SdH} = 1.2\times10^{12}$ cm$^{-2}$ (sample A) and $1.4\times10^{12}$ cm$^{-2}$ (sample B), respectively. These values can be compared to those extracted from the Hall effect, $n_H = (eR_H)^{-1}$, where $R_H$ is the low-field Hall coefficient and $e$ is the elementary charge. We obtain $n_H = 2.1\times10^{12}$ cm$^{-2}$ for sample A and $2.7\times10^{12}$ cm$^{-2}$ for sample B. The presence of bulk carriers, which do not give rise to quantum oscillations, causes $n_H > n_{SdH}$ [1,21]. Residual electrons in the bulk states are due to thermal activation across a small bulk band gap. Figure 1(c) shows the Landau level assignments in the background subtracted $R_{xx}$ data. Only even integer filling factors are observed below 9 T. The two-fold degeneracy is lifted above 9 T for sample A, as in our earlier findings [1].

Figure 2(a) shows $R_{xx}$ from sample A as a function of gate bias ($V_g$) under different applied magnetic fields. As charge carriers are depleted by the application of a negative $V_g$, $R_{xx}$ rapidly increases and reaches the CNP at $V_g = -4.1$ V. At the CNP, the value of $R_{xx}$ is approximately $h/e^2$. With increasing negative $V_g$ beyond the CNP, $R_{xx}$ decreases again, due to the accumulation of holes. At high magnetic fields, SdH oscillations are observed, which we will examine more closely below. Significantly, the value of $R_{xx}$ at the CNP increases by more than an order of magnitude when a magnetic field is applied.

Figure 2(b) shows $|eR_H|^{-1}$ for sample A, which shows the accumulation of $n$-type carriers by an order of magnitude as $V_g$ is swept from negative to positive. As the Fermi level energy is tuned below the CNP, $|eR_H|^{-1}$ increases due to hole accumulation in the two-dimensional states. The combined presence of holes in the surface states and residual electrons in the bulk states compensates $R_H$ (depending on the relative concentrations and mobilities), thereby increasing $|eR_H|^{-1}$. The presence of holes can be clearly seen from the high magnetic field behavior (reversal in slope) of $R_{xy}$, shown in Fig. 2(c). In Fig. 2(d), sample B, which is thicker and has a larger bulk contribution, shows a less pronounced signature of *p*-type carriers, consistent with the interpretation of a background of *n*-type carriers are a residual (thickness-dependent) bulk contribution.

At high magnetic fields, quantum oscillations appear in $R_{xx}$, as shown for sample A in Fig. 3(a). Minima in $R_{xx}$ correspond to integer values of the filling factor $v$. Close to the CNP and at high magnetic field, the two-fold degeneracy of the Landau levels is lifted, as indicated by the superscripts in Fig. 3(a). The maxima shift towards higher $V_g$ at higher magnetic fields, indicating *n*-type transport in the two-dimensional states [Fig. 3(b)]. A contour-plot fan diagram is shown in Fig. 4(a). The color scale reflects $\frac{d^2 \text{Log} R_{xx}}{dV_g^2}$. The minima all extrapolate to the same $V_g$ (-4.1 V), in good agreement to CNP determined by the maximum in $R_{xx}$ and minimum in $|eR_H|^{-1}$ (Fig. 2). Moreover, a minimum, which shifts to lower $V_g$ at higher magnetic fields, can be detected on the negative $V_g$ side of the CNP, which indicates *p*-type transport [see label "-1" in Fig. 4(a)]. This further supports the correct identification of the CNP. One reasonable explanation for the electron-hole asymmetry in the data is a higher effective mass of the holes compared to the electrons. Furthermore, a previous study also detected a greater concentration of defects in the hole states [26].

Figure 4(b) shows the position of the valleys as a function of $n_H$ [extracted from Fig. 2(b)], which also shows the splitting of the levels (2+ and 2-) at 9 T, consistent with Fig. 1(c). In general, the carrier density should follow the relation:

$$n_{SdH} \propto \frac{e}{h} vB, \tag{1}$$

where $v = Ng_v$ is the filling factor, $N$ is the Landau level index, $g_v$ is the degeneracy of the Landau level, and $h$ is Planck's constant. Figure 4(c) shows that all curves in Fig. 4(b) collapse on a single line, when plotted as a function of $vB$, as expected from Eq. (1). This confirms the correct Landau level indexing. Furthermore, it shows that $V_g$ only modulates the carriers that contribute to the SdH effect. This indicates that carriers in the two-dimensional states screen the bulk, as would be expected for carriers in surface states.

To briefly summarize the transport measurements, these clearly indicate that the gate voltage tunes the two-dimensional carrier system in these films through a CNP. As the Fermi level traverses the CNP, signatures of holes are detected in $R_{xy}$ and in the quantum oscillations. These results are a clear indication of a Dirac fermions of the two-dimensional states of this Dirac semimetal.

### B. Quantum capacitance and zero energy Landau level

We can further confirm the Dirac nature of the two-dimensional states by turning to the C-V characteristics, which are shown for sample A in Fig. 5. The quantum capacitance, $C_q$, reflects the density of states ($D$) at the Fermi energy $E_F$, $C_q = D\,e^2$. The CNP appears as a minimum in the capacitance (density of states), which is clearly seen in Fig. 5(a). Furthermore, in a two-dimensional Dirac system, $D$, and thus $C_q$, depend linearly on $E_F$. Here, $C_q$ is in series with the oxide capacitance $C_{ox}$ of the $Al_2O_3$ dielectric, so that the total capacitance is given as $1/C = 1/C_{ox} + 1/C_q$. Figure 5(b) shows the extracted $C_q$ after removal of $C_{ox}$ (0.55 µF/cm$^2$), which was determined from the saturation capacitance, as a function of $\sqrt{n}$ ($\sqrt{p}$), were $n$ and $p$ are the electron and hole carrier densities, respectively. Here, $n$ ($p$) was determined by integration of the C-V curve, assuming that it is zero at the CNP. For a two-dimensional Dirac material, $E_F = \hbar v_F \sqrt{\pi n}$, where $\hbar$ is the reduced Planck's constant and $v_F$ the Fermi velocity. One therefore expects $C_q \propto \sqrt{n}$ (or $C_q \propto \sqrt{p}$), which is what is observed [see also dotted lines in Fig. 5(b)]. The flat $C_q$ near the CNP has been attributed to electron-hole puddles in similar experiments in graphene [27].

The presence of a zero energy Landau level at the CNP is a definite signature of Dirac fermions, but is difficult to detect in transport measurements even in high-quality Dirac materials [27]. The C-V characteristics under an applied magnetic field show quantum oscillations due to the formation of Landau levels (Fig. 5). Moreover, the zero energy Landau level at the CNP can be clearly detected and thus provides additional confirmation of Dirac fermions in the two-dimensional states in these thin films of the three-dimensional Dirac semimetal $Cd_3As_2$. In summary, the C-V measurements clearly detect the CNP, the zero energy Landau level, and the linear dependence of the density of states on the Fermi energy around the CNP, all of which provide evidence for a two-dimensional Dirac electronic spectrum in the confined states of $Cd_3As_2$.

### C. Comparisons with theoretical predictions

Finally, we briefly discuss our results in the context of different theoretical predictions for the surface states in $Cd_3As_2$ and related materials. Scenarios suggested in the theoretical literature are (i) double "Weyl orbits" [10,13,28,29], i.e. the surface Fermi arcs that are connected through the bulk at the Dirac nodes, (ii) independent surface states on bottom and top surfaces [12], similar to a three-dimensional topological insulator, and (iii) the quantum spin Hall insulator state if the Fermi level is at the CNP [4]. Unlike the first two scenarios, the latter is characterized by complete gapping, leaving only the topologically protected one-dimensional edge states in zero magnetic field. Our results show that the gate voltage tunes a single set of Landau levels, which is inconsistent with independent top and bottom surfaces, which should be strongly non-degenerate due to the strong asymmetry of the thin film structure. Clearly, further theoretical and experimental studies are needed, in particular with regards to the nature of the CNP. The observation of the resistance quantum at the CNP could be consistent with a single ballistic edge channel. The rapid increase in $R_{xx}$ under an applied magnetic field when the Fermi level is at the CNP indicates that the magnetic field breaks the protection of this channel from scattering. Non-local transport measurements on much smaller devices, as have been carried out for QSH systems in the literature [30,31], should be employed to probe the nature of the state at the CNP.

## Acknowledgements

The authors thank Jim Allen and Andrea Young for very useful discussions. They also gratefully acknowledge support through by a grant from the U.S. Army Research Office (grant no. W911NF-16-1-0280) and the Vannevar Bush Faculty Fellowship program by the U.S. Department of Defense (grant no. N00014-16-1-2814). The dilution fridge used in the measurements was funded through the Major Research Instrumentation program of the U.S. National Science Foundation (award no. DMR 1531389). This work made use of the MRL Shared Experimental Facilities, which are supported by the MRSEC Program of the U.S. National Science Foundation under Award No. DMR 1720256.

**References**

[1]   T. Schumann, L. Galletti, D. A. Kealhofer, H. Kim, M. Goyal, and S. Stemmer, *Observation of the Quantum Hall Effect in Confined Films of the Three-Dimensional Dirac Semimetal $Cd_3As_2$*, Phys. Rev. Lett. **120**, 016801 (2018).

[2]   C. Zhang, A. Narayan, S. Lu, J. Zhang, H. Zhang, Z. Ni, X. Yuan, Y. Liu, J.-H. Park, E. Zhang, W. Wang, S. Liu, L. Cheng, L. Pi, Z. Sheng, S. Sanvito, and F. Xiu, *Evolution of Weyl orbit and quantum Hall effect in Dirac semimetal $Cd_3As_2$*, Nat. Comm. **8**, 1272 (2017).

[3]   M. Uchida, Y. Nakazawa, S. Nishihaya, K. Akiba, M. Kriener, Y. Kozuka, A. Miyake, Y. Taguchi, M. Tokunaga, N. Nagaosa, Y. Tokura, and M. Kawasaki, *Quantum Hall states observed in thin films of Dirac semimetal $Cd_3As_2$*, Nat. Comm. **8**, 2274 (2017).

[4]   Z. J. Wang, H. M. Weng, Q. S. Wu, X. Dai, and Z. Fang, *Three-dimensional Dirac semimetal and quantum transport in $Cd_3As_2$*, Phys. Rev. B **88**, 125427 (2013).

[5]   Z. K. Liu, J. Jiang, B. Zhou, Z. J. Wang, Y. Zhang, H. M. Weng, D. Prabhakaran, S.-K. Mo, H. Peng, P. Dudin, T. Kim, M. Hoesch, Z. Fang, X. Dai, Z. X. Shen, D. L. Feng, Z. Hussain, and Y. L. Chen, *A stable three-dimensional topological Dirac semimetal $Cd_3As_2$*, Nat. Mater. **13**, 677-681 (2014).

[6]   M. Neupane, S. Y. Xu, R. Sankar, N. Alidoust, G. Bian, C. Liu, I. Belopolski, T. R. Chang, H. T. Jeng, H. Lin, A. Bansil, F. Chou, and M. Z. Hasan, *Observation of a three-dimensional topological Dirac semimetal phase in high-mobility $Cd_3As_2$*, Nat. Comm. **5**, 3786 (2014).

[7] S. Borisenko, Q. Gibson, D. Evtushinsky, V. Zabolotnyy, B. Buchner, and R. J. Cava, *Experimental Realization of a Three-Dimensional Dirac Semimetal*, Phys. Rev. Lett. **113**, 165109 (2014).

[8] X. G. Wan, A. M. Turner, A. Vishwanath, and S. Y. Savrasov, *Topological semimetal and Fermi-arc surface states in the electronic structure of pyrochlore iridates*, Phys. Rev. B **83**, 205101 (2011).

[9] B.-J. Yang and N. Nagaosa, *Classification of stable three-dimensional Dirac semimetals with nontrivial topology*, Nat. Comm. **5**, 4898 (2014).

[10] A. C. Potter, I. Kimchi, and A. Vishwanath, *Quantum oscillations from surface Fermi arcs in Weyl and Dirac semimetals*, Nat. Comm. **5**, 5161 (2014).

[11] E. V. Gorbar, V. A. Miransky, I. A. Shovkovy, and P. O. Sukhachov, *Dirac semimetals $A_3Bi$ (A = Na, K, Rb) as $Z_2$ Weyl semimetals*, Phys. Rev. B **91**, 121101 (2015).

[12] M. Kargarian, M. Randeria, and Y. M. Lu, *Are the surface Fermi arcs in Dirac semimetals topologically protected?*, Proc. Natl. Acad. Sci. **113**, 8648-8652 (2016).

[13] J. W. Villanova, E. Barnes, and K. Park, *Engineering and Probing Topological Properties of Dirac Semimetal Films by Asymmetric Charge Transfer*, Nano Lett. **17**, 963-972 (2017).

[14] S. M. Young, S. Zaheer, J. C. Y. Teo, C. L. Kane, E. J. Mele, and A. M. Rappe, *Dirac Semimetal in Three Dimensions*, Phys. Rev. Lett. **108**, 140405 (2012).

[15] C. X. Liu, H. Zhang, B. H. Yan, X. L. Qi, T. Frauenheim, X. Dai, Z. Fang, and S. C. Zhang, *Oscillatory crossover from two-dimensional to three-dimensional topological insulators*, Phys. Rev. B **81**, 041307 (2010).

[16] A. Narayan, D. Di Sante, S. Picozzi, and S. Sanvito, *Topological Tuning in Three-Dimensional Dirac Semimetals*, Phys. Rev. Lett. **113**, 256403 (2014).

[17] A. Akrap, M. Hakl, S. Tchoumakov, I. Crassee, J. Kuba, M. O. Goerbig, C. C. Homes, O. Caha, J. Novak, F. Teppe, W. Desrat, S. Koohpayeh, L. Wu, N. P. Armitage, A. Nateprov, E. Arushanov, Q. D. Gibson, R. J. Cava, D. van der Marel, B. A. Piot, C. Faugeras, G. Martinez, M. Potemski, and M. Orlita, *Magneto-Optical Signature of Massless Kane Electrons in $Cd_3As_2$*, Phys. Rev. Lett. **117**, 136401 (2016).

[18] S. Murakami, *Phase transition between the quantum spin Hall and insulator phases in 3D: emergence of a topological gapless phase*, New J. Phys. **9**, 356 (2007).

[19] H. Pan, M. M. Wu, Y. Liu, and S. A. Yang, *Electric control of topological phase transitions in Dirac semimetal thin films*, Sci. Rep. **5**, 14639 (2015).

[20] X. Xiao, S. A. Yang, Z. Liu, H. Li, and G. Zhou, *Anisotropic Quantum Confinement Effect and Electric Control of Surface States in Dirac Semimetal Nanostructures*, Sci. Rep. **5**, 7898 (2015).

[21] M. Goyal, L. Galletti, S. Salmani-Rezaie, T. Schumann, D. A. Kealhofer, and S. Stemmer, *Thickness dependence of the quantum Hall effect in films of the three-dimensional Dirac semimetal $Cd_3As_2$*, APL Mater. **6**, 026105 (2018).

[22] T. Schumann, M. Goyal, H. Kim, and S. Stemmer, *Molecular beam epitaxy of $Cd_3As_2$ on a III-V substrate*, APL Mater. **4**, 126110 (2016).

[23] M. N. Ali, Q. Gibson, S. Jeon, B. B. Zhou, A. Yazdani, and R. J. Cava, *The Crystal and Electronic Structures of $Cd_3As_2$, the Three-Dimensional Electronic Analogue of Graphene*, Inorg. Chem. **53**, 4062−4067 (2014).

[24]   See Supplemental Material [link to be inserted by publisher] for magnetoresistance measurements on samples without gate dielectrics and for a comparison of the magnetoresistance oscillations for different sweep directions of the gate voltage.

[25]   W. E. Spicer, P. W. Chye, P. R. Skeath, C. Y. Su, and I. Lindau, *New and Unified Model for Schottky-Barrier and Iii-v Insulator Interface States Formation*, J.Vac. Sci. & Technol. **16**, 1422-1433 (1979).

[26]   S. Jeon, B. B. Zhou, A. Gyenis, B. E. Feldman, I. Kimchi, A. C. Potter, Q. D. Gibson, R. J. Cava, A. Vishwanath, and A. Yazdani, *Landau quantization and quasiparticle interference in the three-dimensional Dirac semimetal $Cd_3As_2$*, Nat. Mater. **13**, 851-856 (2014).

[27]   L. A. Ponomarenko, R. Yang, R. V. Gorbachev, P. Blake, A. S. Mayorov, K. S. Novoselov, M. I. Katsnelson, and A. K. Geim, *Density of States and Zero Landau Level Probed through Capacitance of Graphene*, Phys. Rev. Lett. **105**, 136801 (2010).

[28]   P. J. W. Moll, N. L. Nair, T. Helm, A. C. Potter, I. Kimchi, A. Vishwanath, and J. G. Analytis, *Transport evidence for Fermi-arc-mediated chirality transfer in the Dirac semimetal $Cd_3As_2$*, Nature **535**, 266-270 (2016).

[29]   C. M. Wang, H.-P. Sun, H.-Z. Lu, and X. C. Xie, *3D Quantum Hall Effect of Fermi Arc in Topological Semimetals*, Phys. Rev. Lett. **119**, 136806 (2017).

[30]   A. F. Young, J. D. Sanchez-Yamagishi, B. Hunt, S. H. Choi, K. Watanabe, T. Taniguchi, R. C. Ashoori, and P. Jarillo-Herrero, *Tunable symmetry breaking and helical edge transport in a graphene quantum spin Hall state*, Nature **505**, 528-532 (2014).

[31]  A. Roth, C. Brune, H. Buhmann, L. W. Molenkamp, J. Maciejko, X. L. Qi, and S. C. Zhang, *Nonlocal Transport in the Quantum Spin Hall State*, Science **325**, 294-297 (2009).

# Figure Captions

**Figure 1:** Magnetoresistance and quantum Hall effect of samples A and B (with gate dielectric). (a) $R_{xx}$ and $R_{xy}$ as a function of magnetic field, showing SdH oscillations in $R_{xx}$ and incipient quantum Hall plateaus in $R_{xy}$. (b) Landau fan diagrams (c) SdH oscillations after background subtraction. Lifting of the two-fold degeneracy can be seen above 9 T for Sample A but not sample B.

**Figure 2:** Magnetotransport behavior of the gated devices. (a) $R_{xx}$ of sample A as a function of gate bias ($V_g$) for different magnetic fields. (b) $|eR_H|^{-1}$ extracted from the low-field Hall effect as a function of $V_g$ for sample A. (c-d) $R_{xy}$ of samples A and B, respectively, as a function of the magnetic field, for large negative values of $V_g$ (Fermi level below the CNP of the two-dimensional states). Data were anti-symmetrized using the relation $(R_{xy}(B) - R_{xy}(-B))/2$, to suppress contributions from the noise and intermixing of $R_{xx}$ and $R_{xy}$ signals. The inset in (b) shows a schematic of the gated Hall bar structure (bottom) and sketches that illustrate the tuning of the Fermi level (dashed line) with respect to the two-dimensional states (blue) at different $V_g$. Bulk states are shown in gold.

**Figure 3:** SdH oscillations as a function of $V_g$ for sample A. (a) $R_{xx}$ as a function of $V_g$ at 14 T. Landau level indices of the minima are indicated. (b) Magnetic field behavior of the SdH oscillations, offset for clarity. Data were symmetrized using the relation $(R_{xx}(B) + R_{xx}(-B))/2$.

**Figure 4:** Fan diagram. (a) Plot of $\frac{d^2 \text{Log} R_{xx}}{dV_g^2}$ (arbitrary units) as a function of $B$ and $V_g$ for sample A. A smoothing procedure prior to each differentiation was performed to increase the signal to noise ratio. (b) Position of the maxima in (a), as extracted with a peak finding algorithm, as a function of the carrier density $n_H$, determined from the Hall effect, as shown in Fig. 2. (c) Same data as in (b), but plotted as $\nu B$, where $\nu$ is the filling factor, which collapse on a single straight line. Labels in (a) and (b) indicate the Landau level $N$.

**Figure 5:** C-V characteristics. (a) Capacitance of sample A measured as a function of gate voltage with and without an applied magnetic field of 14 T. (b) Quantum capacitance extracted from the data shown in (a), as a function of the square root of the modulated carrier density. The dotted lines are a fit to the data at zero magnetic field.

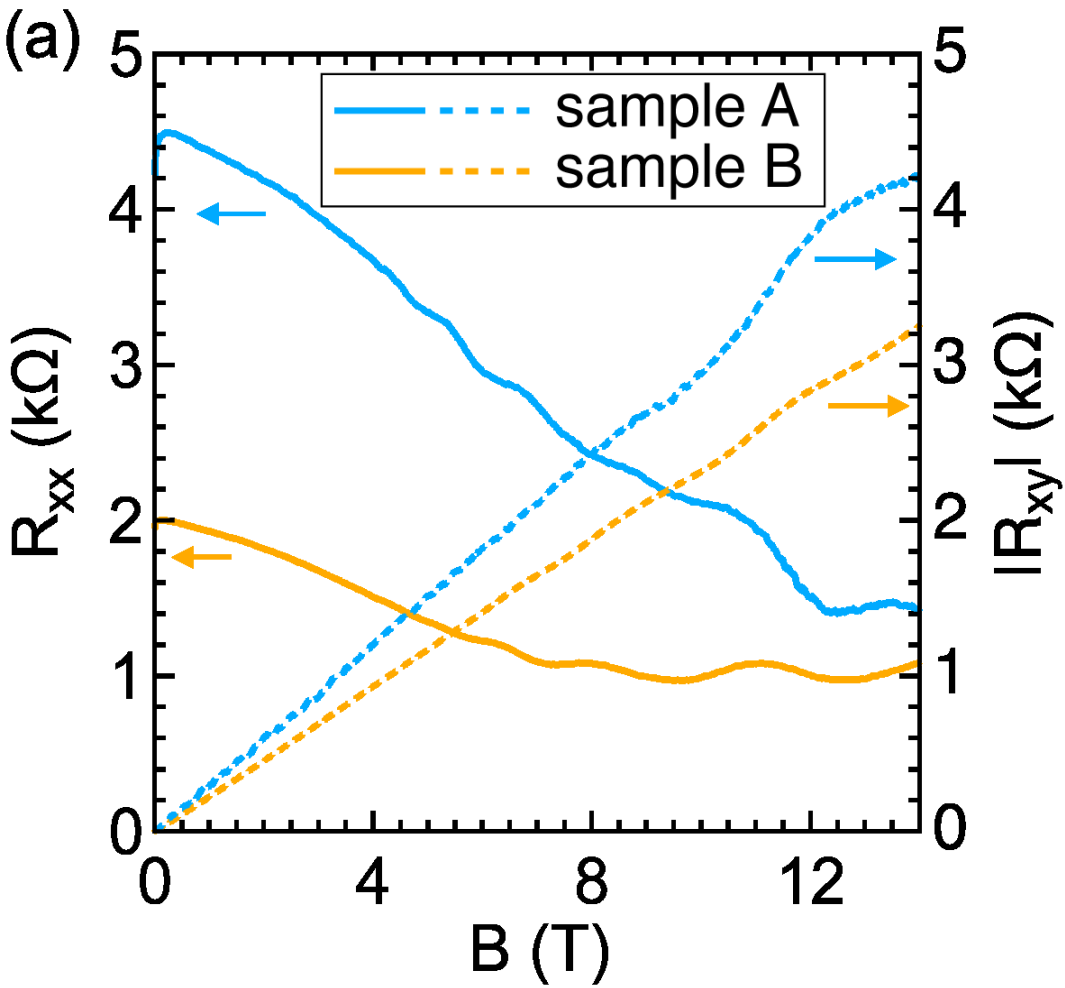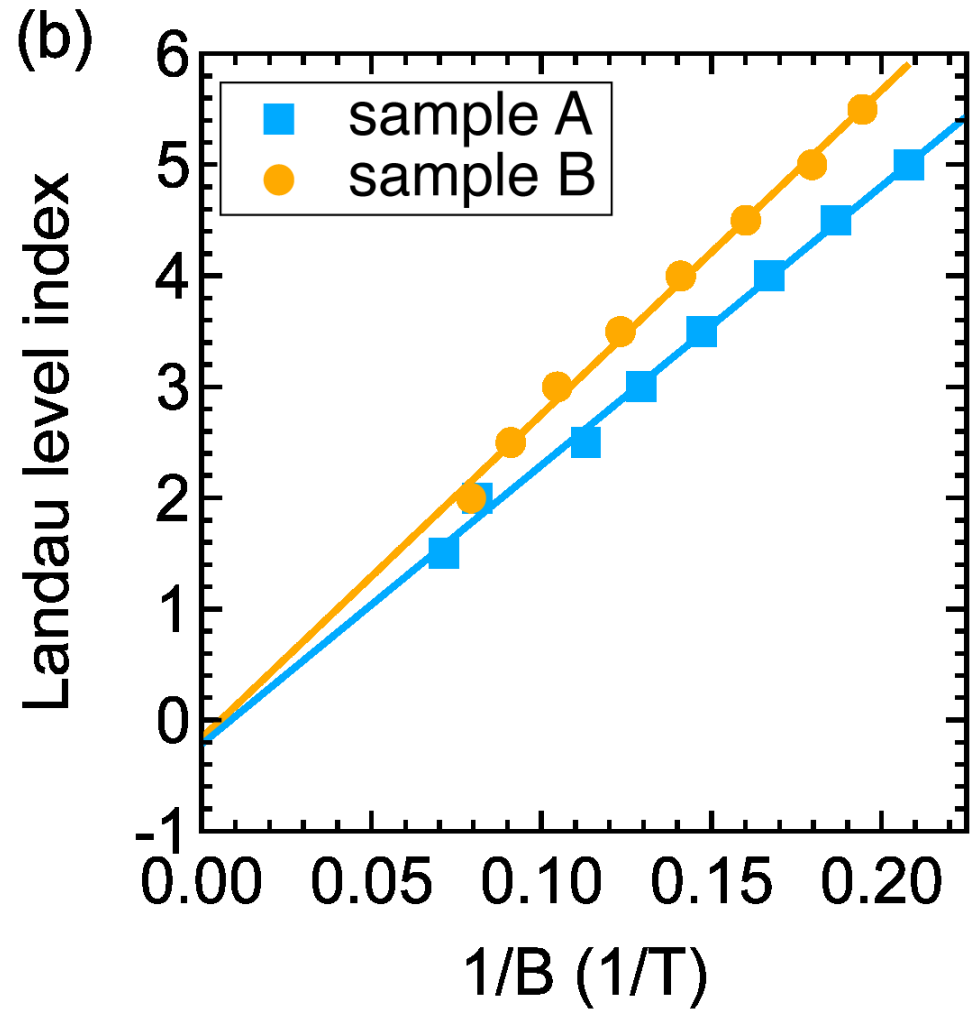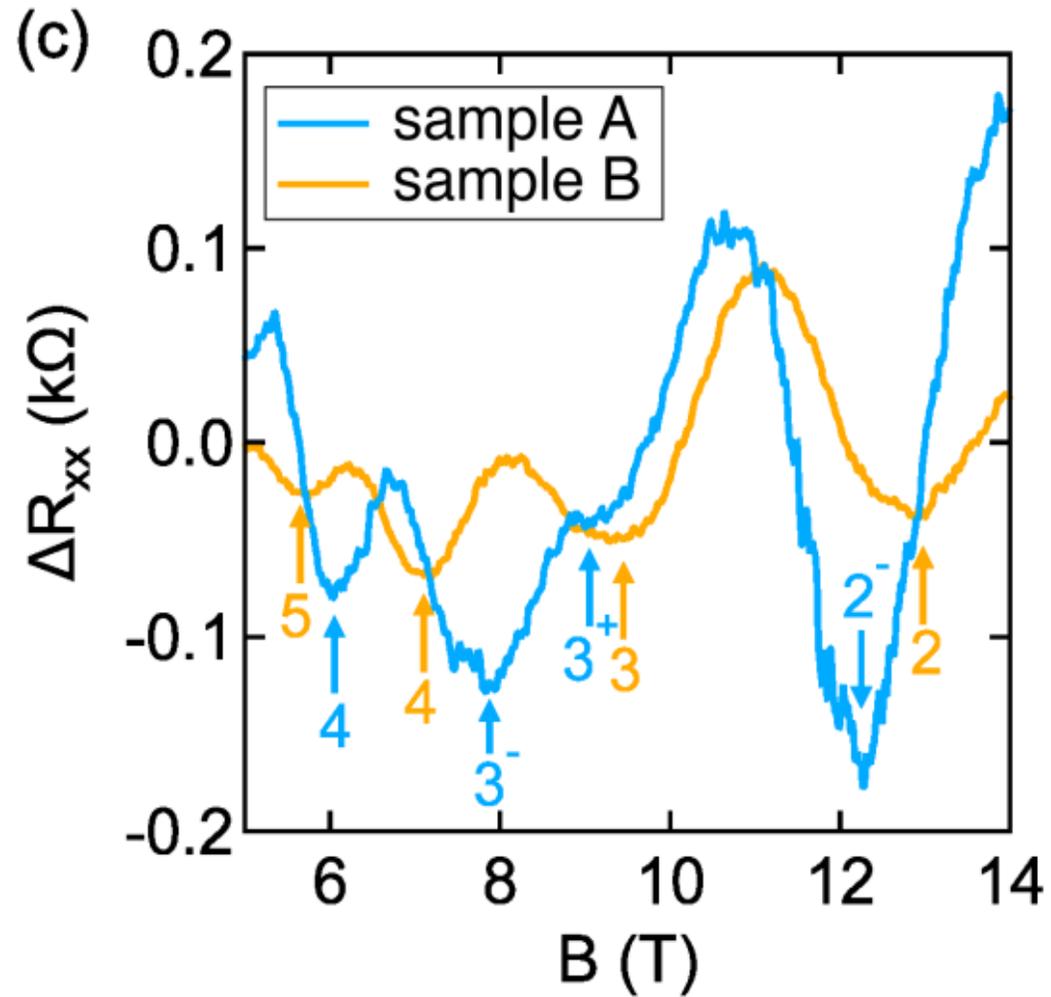

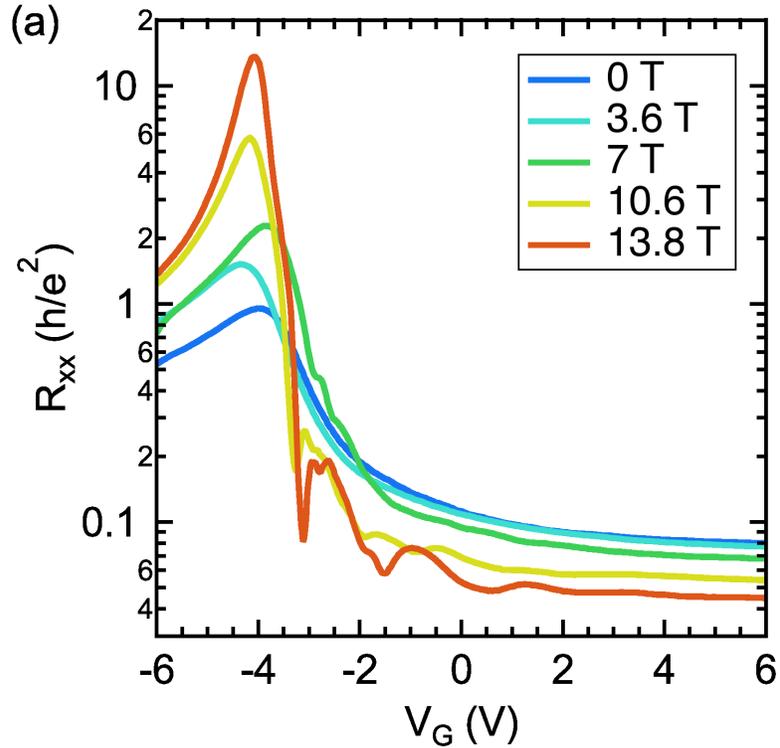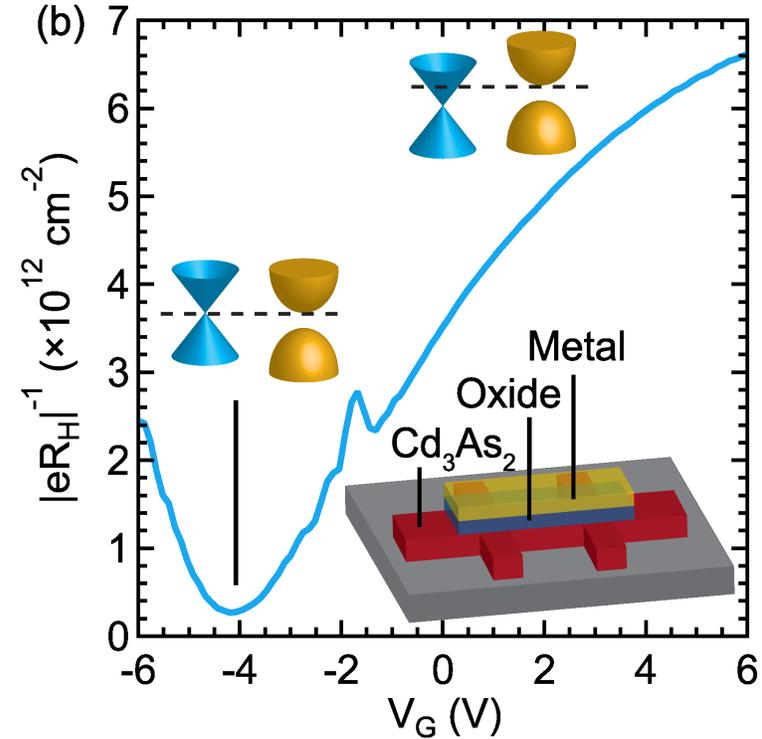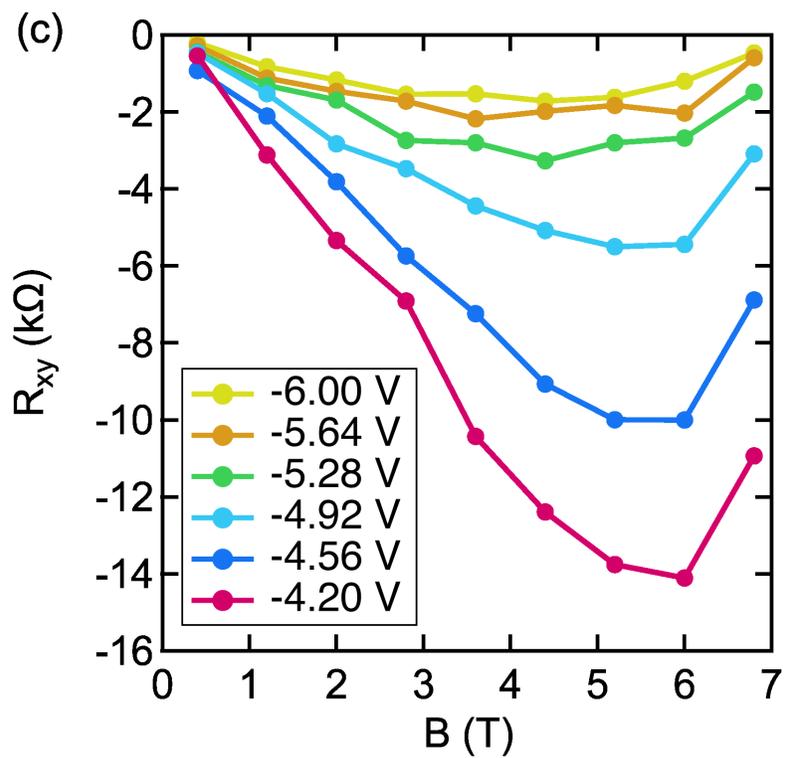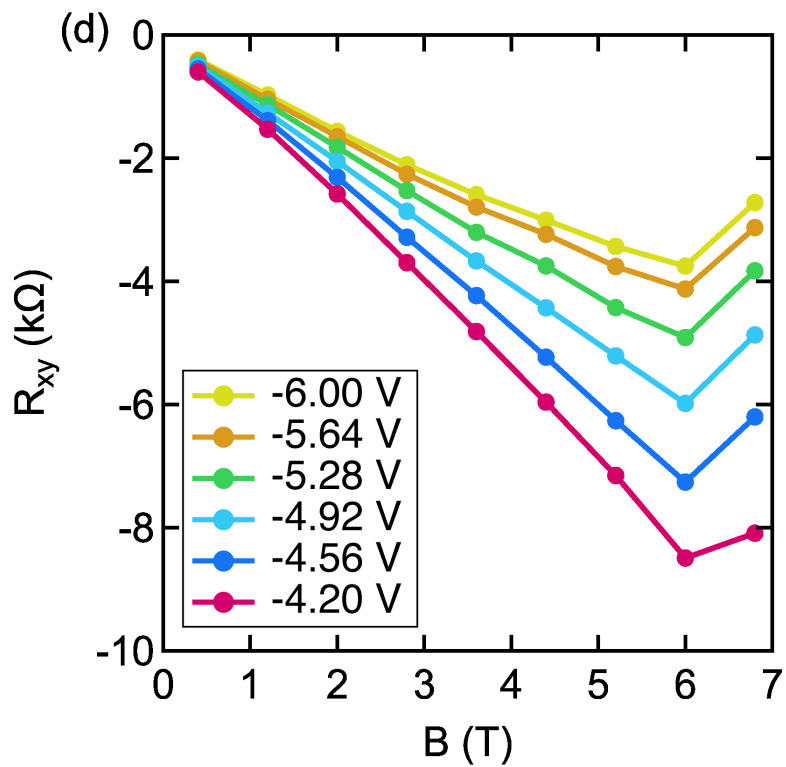

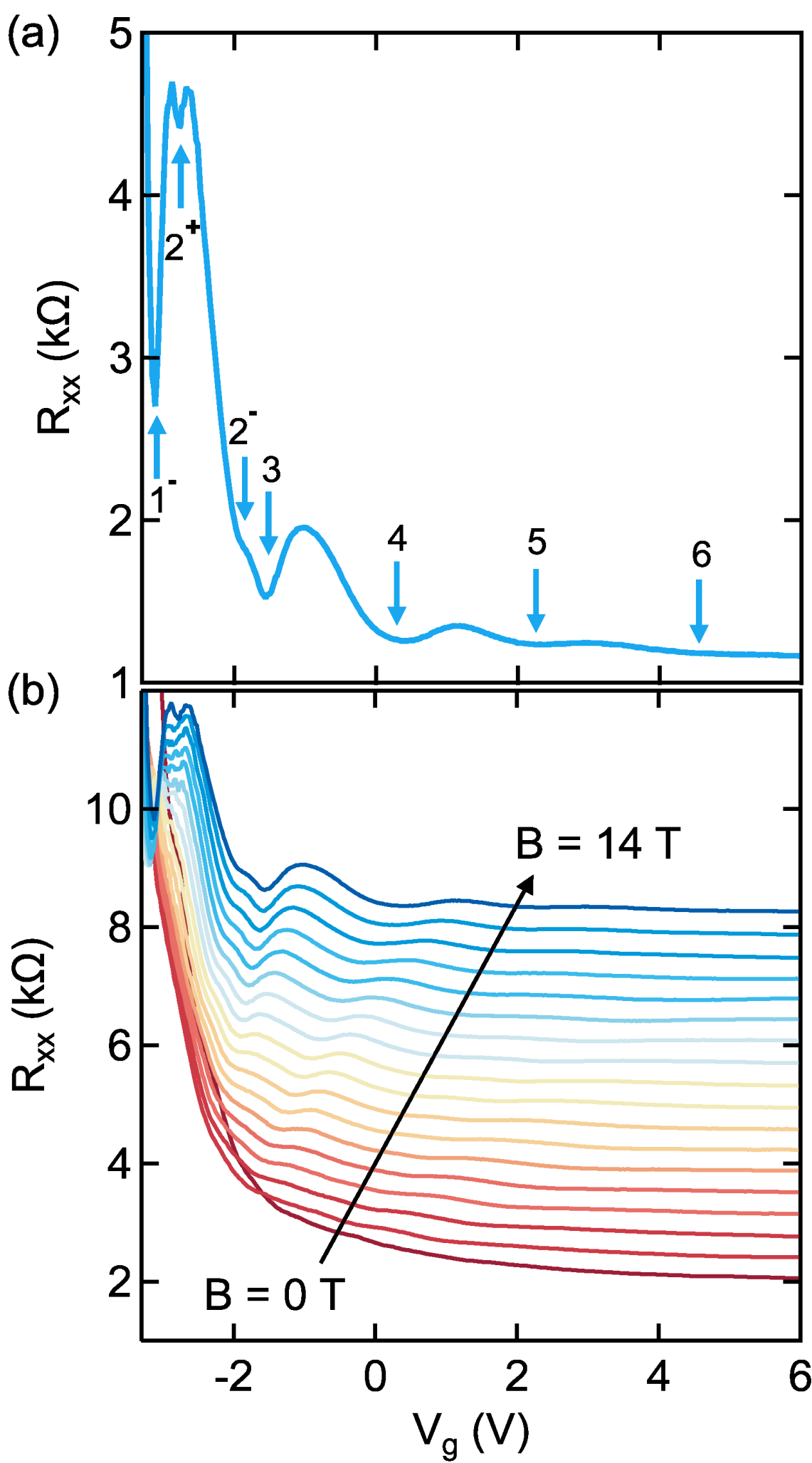

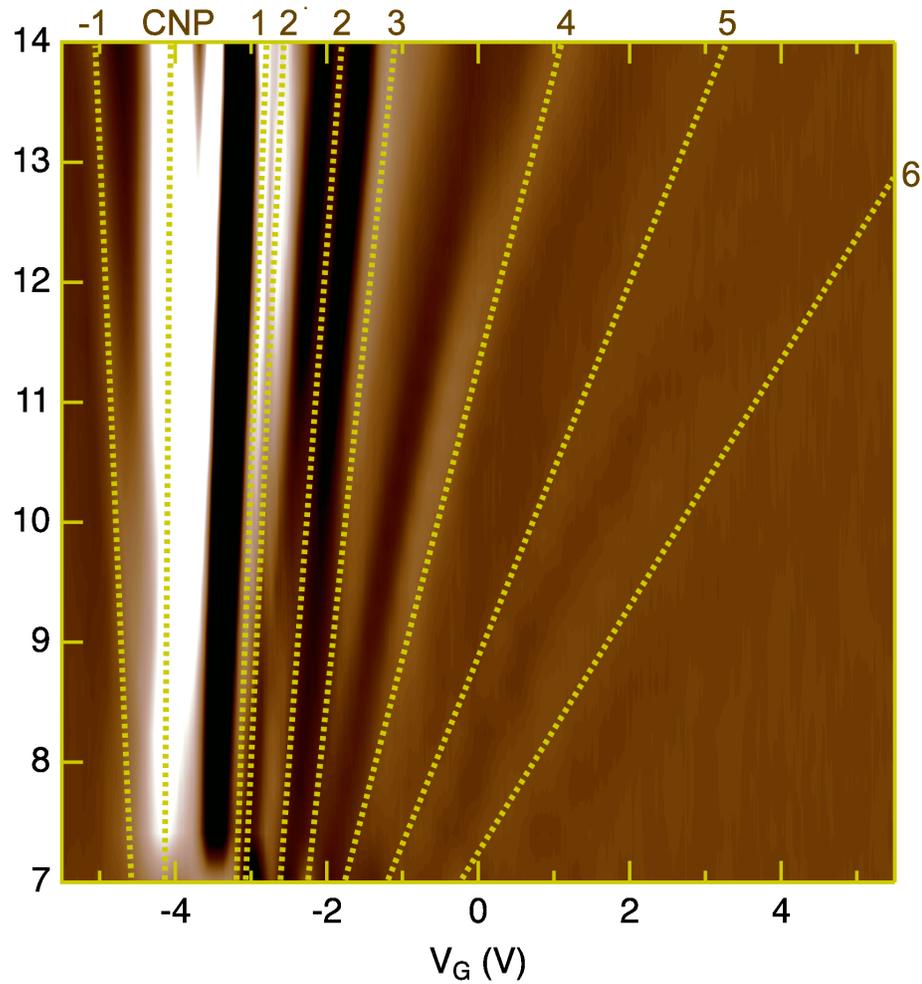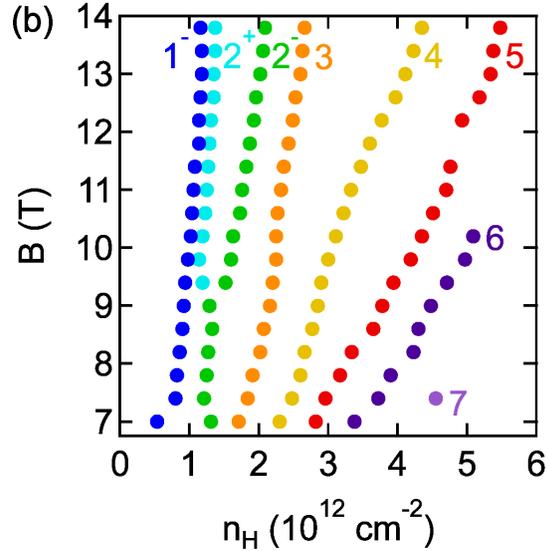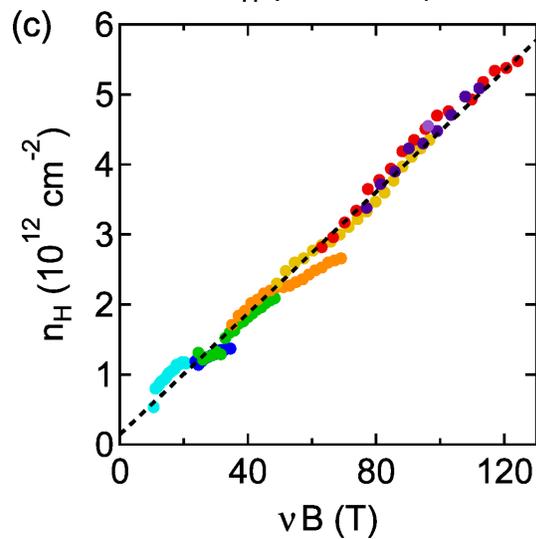

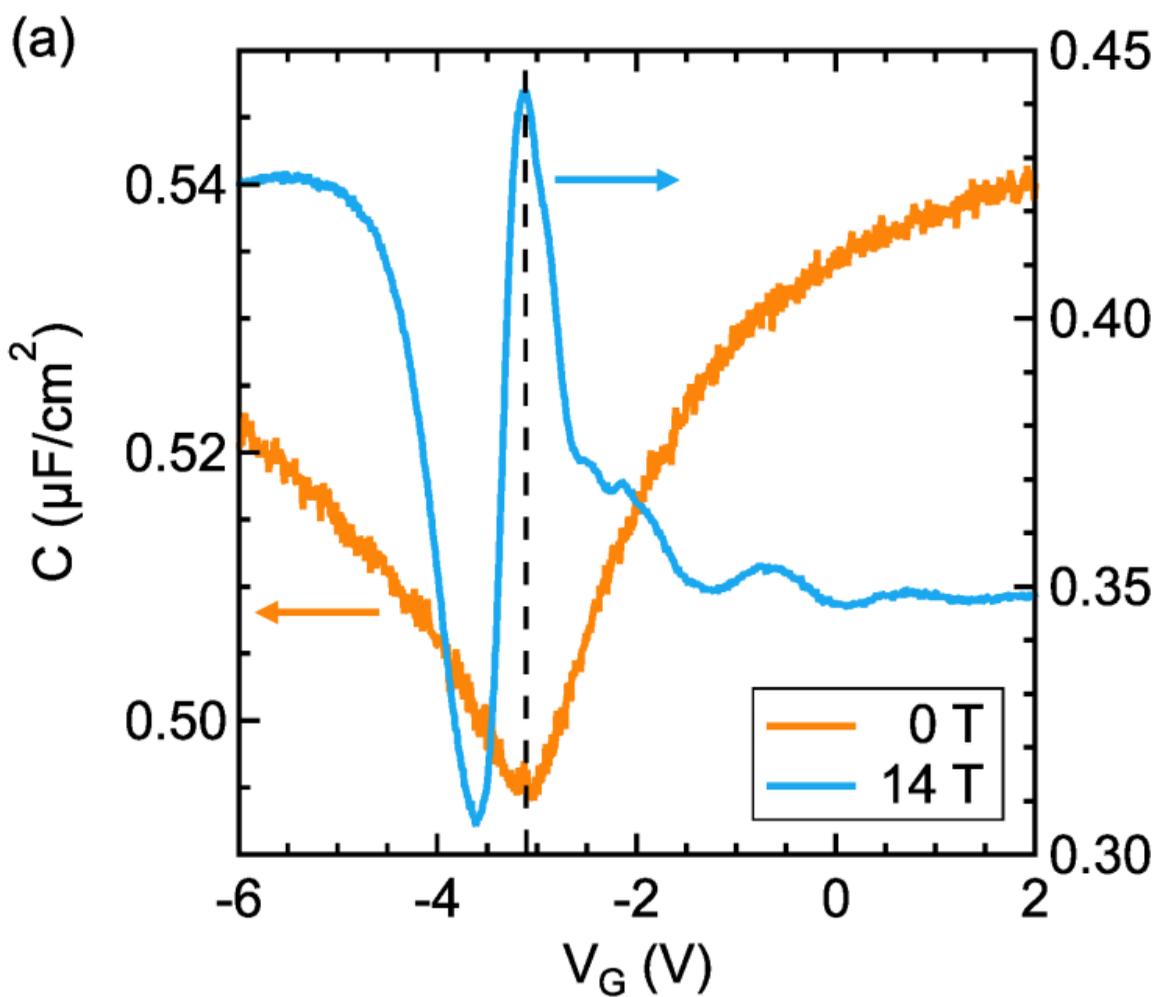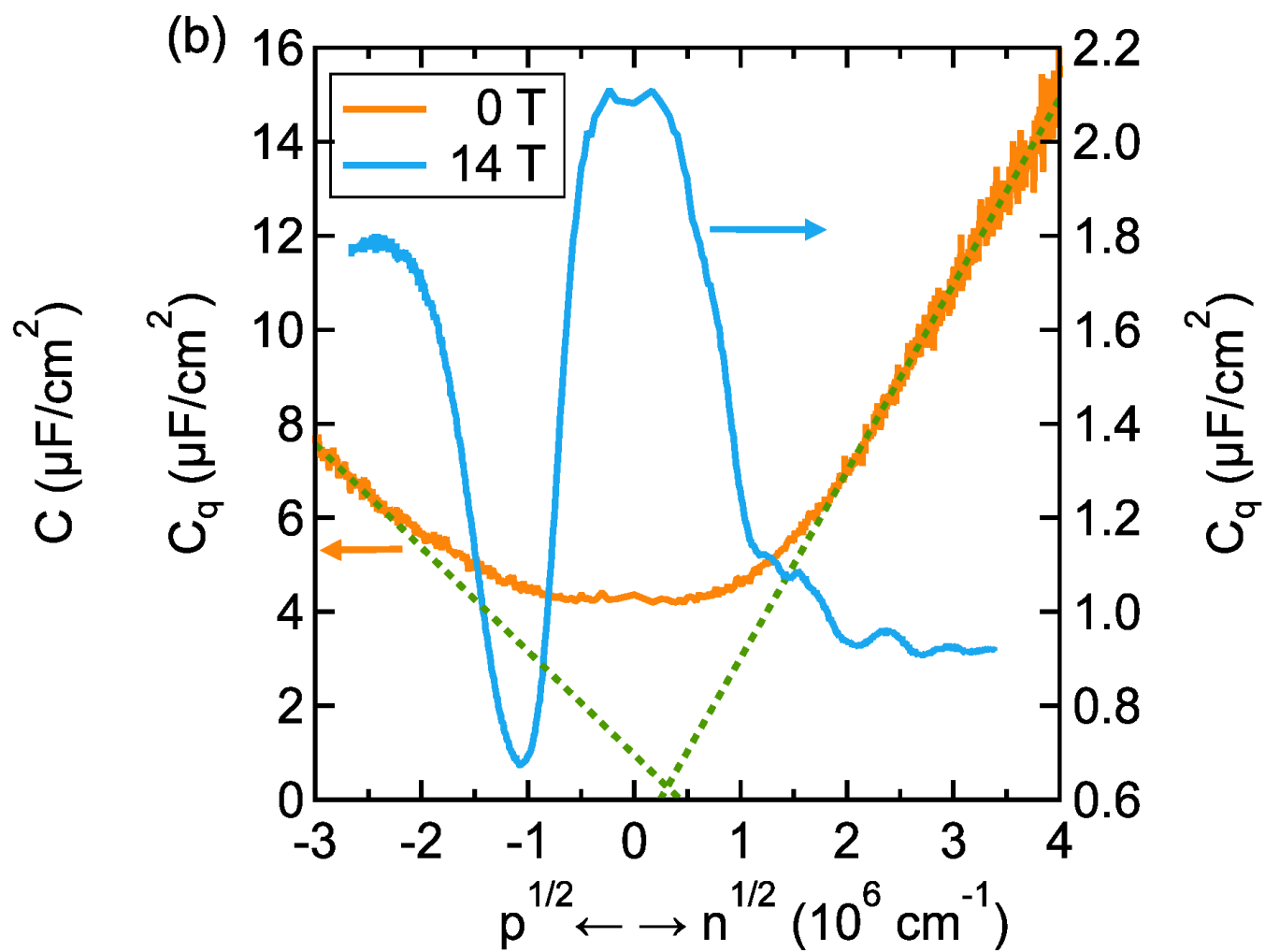